%% file: main.tex
\documentclass[11pt]{article}

\usepackage[preprint]{acl}
\usepackage{xurl}

\usepackage{times}
\usepackage[T1]{fontenc}
\usepackage[utf8]{inputenc}
\usepackage{microtype}
\usepackage{inconsolata}
\usepackage[hang,flushmargin]{footmisc}

\usepackage{xspace}
\usepackage{enumitem}
\usepackage{tabularx}
\usepackage{amssymb}
\usepackage{amsmath}
\usepackage{amsfonts}
\usepackage{mathtools}
\usepackage{graphicx}
\usepackage{booktabs}
\usepackage{listings}
\usepackage{fancyvrb}
\usepackage{tikz}
\usepackage{tikz-cd}
\usepackage{pgf}
\usepackage{pgfplots}
\usepackage{subcaption}
\usepackage{float}
\usepackage{adjustbox}
\usepackage{xcolor}
\usepackage{algorithm}
\usepackage{algpseudocode}
\usepackage{tcolorbox}
\usepackage{mdframed}
\usepackage{verbatim}
\usepackage{ulem}

\usetikzlibrary{arrows.meta,positioning,fit,shapes,calc,backgrounds}

\pgfdeclarelayer{background}
\pgfsetlayers{background,main}

\usetikzlibrary{shapes,arrows,positioning}
\tcbuselibrary{listingsutf8, skins, breakable}

\tcolorboxenvironment{SemanticBox}{
  enhanced,
  colback=blue!5!white,
  colframe=blue!40!black,
  sharp corners,
  boxrule=0.5pt,
  before skip=4pt,
  after skip=4pt,
  breakable
}

\tcolorboxenvironment{RawBox}{
  enhanced,
  colback=orange!5!white,
  colframe=orange!40!black,
  sharp corners,
  boxrule=0.5pt,
  before skip=4pt,
  after skip=4pt,
  breakable
}

\newtcolorbox{rqbox}{
  colback=gray!10,
  colframe=black,
  boxrule=0.5pt,
  arc=5pt,
  left=12pt,
  right=12pt,
  top=6pt,
  bottom=6pt,
  before skip=8pt,
  after skip=8pt
}

\input{macros}

\title{Bridging Search Intent and Tabular Data in Dataset Search}

\author{Wenjing Lin \\
  University of California, Berkeley \\
  \texttt{wenjing.lin@berkeley.edu} \\\And
  Madelon Hulsebos \\
  Centrum Wiskunde \& Informatica \\
  \texttt{madelon@cwi.nl} \\}

\begin{document}
\maketitle

\begin{abstract}
Dataset search remains a bottleneck in data analysis workflows, as existing systems largely rely on keyword matching or shallow retrieval that fail to capture user intent. To facilitate intent-driven search, we present two retrieval components to equip multi-faceted dataset search interfaces through natural language. First, we propose Hypothetical Schema Embeddings (HySE) that uses a language model to generate task-aligned hypothetical table schemas which are used for embedding-based retrieval, bridging the gap between intent and table content. We introduce \dataset, a corpus comprising 6,503 tables and diverse task-queries derived from Kaggle, and show that HySE consistently improves recall over strong keyword and embedding-based baselines, as well as over a dense table retriever fine-tuned on \dataset. We also show that HySE generalizes to end-to-end question answering on, for example, the NQ-Tables dataset, a corpus of 169,898 Wikipedia tables with human-written queries, where HySE improves Recall@10 by +6.0\,pp over dense retrieval baselines. Second, we introduce FlexPrune, which translates dataset constraints expressed in natural-language into SQL predicates over a customizable metadata index, enabling filtering across arbitrary metadata attributes. Combined with HySE, FlexPrune increases Recall@10 by +6.0\,pp over dense retrieval followed by tag-based filtering. Our results show that natural-language-based retrieval supporting \textit{task} and metadata queries offers a viable alternative to schema-level keyword search in dataset search.
\end{abstract}

\input{sections/introduction}

\input{sections/related_work}

\input{sections/design_considerations}
\input{sections/system_proposal}

\input{sections/hyse}

\input{sections/flexprune}

\input{sections/sys_implement_condensed}
\input{sections/dataset_construction}

\input{sections/sys_eval_condensed}

\input{sections/conclusion}

\input{sections/limitations}

\section*{Use of Generative AI}
We disclose all use of generative AI in this work:

\paragraph{GenAI as part of the proposed method and dataset.} Large language
models are an integral and studied component of the contributions of this
paper. Specifically: (i)~HySE uses an LLM to generate task-aligned
hypothetical table schemas; (ii)~FlexPrune uses an LLM to translate
natural-language constraints into SQL predicates; and (iii)~the
construction of the \dataset{} corpus uses an LLM (GPT-4o) for metadata
enhancement, inferring temporal- and geographic-granularity attributes for
tables, and to generate the task, keyword, and metadata queries. These uses are described in detail in the corresponding sections and constitute part of the technical contribution of the paper.

\paragraph{GenAI for coding and writing assistance.} Generative AI tools were
used to assist the authors in preparing this work: (i)~to assist with
writing portions of the system implementation and evaluation code; and
(ii)~for editing and proofreading the manuscript.
GenAI was not used to inform the research ideas, system design, evaluation
protocol, or the analysis and interpretation of experimental results. All
AI-assisted code and text were reviewed and verified by the authors, who
take full responsibility for the content of this paper.

\bibliography{references}

\appendix

\input{appendix/prompts}
\input{appendix/kaggle_example}
\input{appendix/ablation_analysis}

\end{document}

%% file: macros.tex
\newcommand{\dataset}{\textsc{KaggleDS}\xspace}

\newcommand{\boldpar}[1]{%
  \par\noindent\textbf{#1}%
}

%% file: sections/introduction.tex
\section{Introduction}
\label{sec:intro}

Dataset search is a recurring bottleneck in data analysis~\cite{Hulsebos2024HILDA}. As data lakes and public repositories grow, analysts spend considerable effort locating relevant datasets before analysis can begin. Industry surveys report that data scientists spend roughly 25\% of their time searching for datasets or preparing data for analysis \cite{Grover2019Amundsen}, and data preparation remains the most time-consuming stage of the workflow according to a survey of more than 2,000 practitioners \cite{Anaconda2023DSReport}. These observations highlight the need for more effective dataset search systems.

\textbf{In Practice, Dataset Search is Still Primarily Keyword-based.}
In enterprise environments, data catalogs such as Amundsen and DataHub store metadata describing datasets and provide search interfaces over these metadata repositories \cite{Grover2019Amundsen,Datahub2025}. At web scale, systems such as Google Dataset Search index datasets using supplied metadata to support keyword lookup \cite{Brickley2019GoogleSearch,Sostek2024GoogleDatasetSearch}. In both settings, retrieval is driven primarily by matching query keywords to metadata fields such as dataset titles, descriptions, or column names. This approach is effective when users already know the relevant dataset names or metadata attributes. However, users may have a clear analytical objective but are unfamiliar with the dataset search space, let alone the exact keywords matching the indexed metadata.

\textbf{Dataset Search Beyond Data Augmentation.}
Prior research on dataset search has largely focused on identifying structural relationships between tables, such as join discovery, schema matching, and table union search \cite{Nargesian2018TableUnionSearch,Zhu2019JOSIE,Zhu2016LSHEnsemble, Fernandez2018Aurum,Srinivas2023LakeBench}. These approaches retrieve tables that can be combined with a known reference table or schema. However, dataset search in practice typically starts from an analytical objective rather than a known table or schema. User studies suggest that analysts reason in terms of analysis goals during early exploration, rather than in terms of specific table schemas \cite{Paton2023Survey}. And an analyst who wants to \emph{“analyze trends in California real estate over the past decade”} may not know upfront which data attributes are needed and how relevant datasets are structured. %

\textbf{Design Requirements from User Research.} In this work, we build on a survey of 89 data professionals about their dataset search challenges and requirements~\cite{Hulsebos2024HILDA}. The survey identified four design desiderata for dataset search systems: (1) \textit{task-driven search} that starts from analytical objectives rather than dataset specifications, (2) \textit{hybrid querying} that combines semantic table embeddings with rich metadata filtering, (3) \textit{result comprehensibility and diversity} to reduce cognitive overload in large catalogs, and (4) \textit{iterative interfaces} that support multi-step refinement.

\textbf{Task-Driven Table Retrieval and Flexible Metadata Pruning via Natural Language.}
We explore how dataset search systems can move beyond keyword- and schema-bound interfaces toward \emph{task-driven, open-ended natural-language interaction}.
Effective retrieval in dataset search builds on dataset semantics as well as metadata \cite{Hearst2006FacetedSearch,Hulsebos2024HILDA}, which we address through two distinct components. First, we introduce \emph{Hypothetical Schema Embeddings (HySE)}, a semantic retrieval mechanism that uses a language model to translate a user’s analytical task description into one or more task-aligned \textit{hypothetical} table schemas, which are then used as inputs for dense retrieval.
Second, we introduce FlexPrune, an attribute-agnostic metadata filter stage that parses natural-language constraints into structured SQL predicates over a customizable metadata index.
Together, these components enable task-driven dataset search in which users can refine their search using natural language.

\textbf{Contributions.} In this work we make the following contributions:
\noindent(1) We propose \textbf{Hypothetical Schema Embeddings (HySE)}, an algorithm that turns a natural-language task into one or more hypothetical schemas for dense retrieval, to bridge the intent-dataset gap.
\noindent(2) We propose \textbf{FlexPrune}, an attribute-agnostic natural-language mechanism for search over metadata. FlexPrune translates NL constraints into SQL over a hybrid metadata index. %
\noindent(3) We present \textbf{\textsc{KaggleDS}}, a new task-based dataset search corpus comprising 6{,}503 tables from 3{,}141 real-world Kaggle datasets, with task-oriented queries in natural language for training and evaluation.
\noindent(4) We present an \textbf{evaluation} across key table retrieval paradigms which shows how HySE improves top-$k$ recall---over keyword and dense retrieval baselines as well as a dense table retriever trained on our corpus---while in combination with FlexPrune yields a significant reduction in result set size.

%% file: sections/related_work.tex
\section{Related Work}
\label{sec:related-work}

\noindent\textbf{Keyword-Based Search.}\ \ Early approaches to dataset discovery relied on keyword search over metadata and content. Early work by \citet{Cafarella2008WebTables} introduced WebTables, which extracted 14.1 billion HTML tables from Google's web crawl and developed schema-aware ranking algorithms that consider table structure and attribute relationships. Google Dataset Search \cite{Brickley2019GoogleSearch} later provided the first large-scale commercial implementation using schema.org markup for web-scale dataset discovery. Within enterprise environments, systems like Amundsen \cite{Grover2019Amundsen} and DataHub \cite{Datahub2025} focus on metadata search with usage-based ranking and lineage tracking. Auctus \cite{Castelo2021Auctus} extends traditional keyword search by supporting spatial, temporal, and data integration queries that enable users to find join-compatible datasets.

\noindent\textbf{Semantic and Embedding-Based Search.}\ \  Subsequent work applies pre-trained language models to move beyond lexical matching toward semantic similarity. Recent approaches include CHORUS \cite{Kayali2024CHORUS}, which unifies multiple data discovery tasks using large language models, and Starmie \cite{Fan2023Starmie}, which employs contrastive learning with pre-trained language models to achieve contextualized column embeddings. TableQA systems like TAPAS \cite{Herzig2020TAPAS} extend BERT \cite{Devlin2019Bert} to encode tabular input, demonstrating that language models can represent table structure. Building on TAPAS, the Dense Table Retriever (DTR) \cite{Herzig2021DTR} trains a specialized encoder to score a natural-language question against a flattened table. DTR is the closest learned counterpart to our setting, and we therefore adopt it as a trained retrieval baseline. SOLO \cite{Wang2023SOLO} addresses natural-language table retrieval via self-supervised learning without requiring manual training data. %

\noindent\textbf{Benchmarks for Dataset Search and Retrieval.} \ \
Several datasets have been used to evaluate dataset discovery systems.
Early studies relied on large web table collections such as WebTables \cite{Cafarella2008WebTables}. Subsequent work constructed data lake benchmarks from open data portals and enterprise repositories~\cite{Khatiwada2022SANTOS, Fan2023Starmie, Fernandez2018Aurum}, as well as union-search benchmarks over open data and data lakes~\cite{Nargesian2018TableUnionSearch, Srinivas2023LakeBench}. System-oriented work such as Voyager \cite{Bogatu2022Voyager} evaluates discovery over small, domain-specific collections.
However, no dataset exists that enables training and evaluating task-based dataset search. To this end, we contribute \dataset, a new dataset derived from thousands of real-world Kaggle datasets with task-oriented queries.

%% file: sections/system_proposal.tex
\section{Components for Bridging Search Intent and Tabular Datasets}
\label{sec:proposal}

Here, we introduce two core components for task-driven multi-faceted dataset search based on natural language interfaces, covering both over dataset content and metadata.
First, \emph{HySE} (§~\ref{sec:hyse}) performs task-based retrieval: given a natural-language task description, it generates hypothetical table schemas and uses their embeddings to retrieve a broad candidate set via ANN search. Second, \emph{FlexPrune} (§~\ref{sec:nl-refine}) can narrow retrieved results down through NL-based metadata filtering, translating user constraints into SQL predicates over an extensible metadata index.

%% file: sections/hyse.tex
\subsection{HySE: Task-based Semantic Search}%
\label{sec:hyse}

To address the \emph{task-driven} desideratum, we introduce \emph{Hypothetical Schema Embeddings (HySE)}, a retrieval mechanism that enables dataset search using task-queries which have less semantic resemblance with the target schemas than user-informed keywords. HySE builds on the classical idea of query expansion \cite{Salton1971QueryExpansion} and draws direct inspiration from hypothetical document embeddings \cite{Gao2022HyDE}, a framework that embeds LLM-generated hypothetical documents to bridge the semantic gap between input query and documents.%

\begin{figure}
    \centering
    \includegraphics[width=\linewidth]{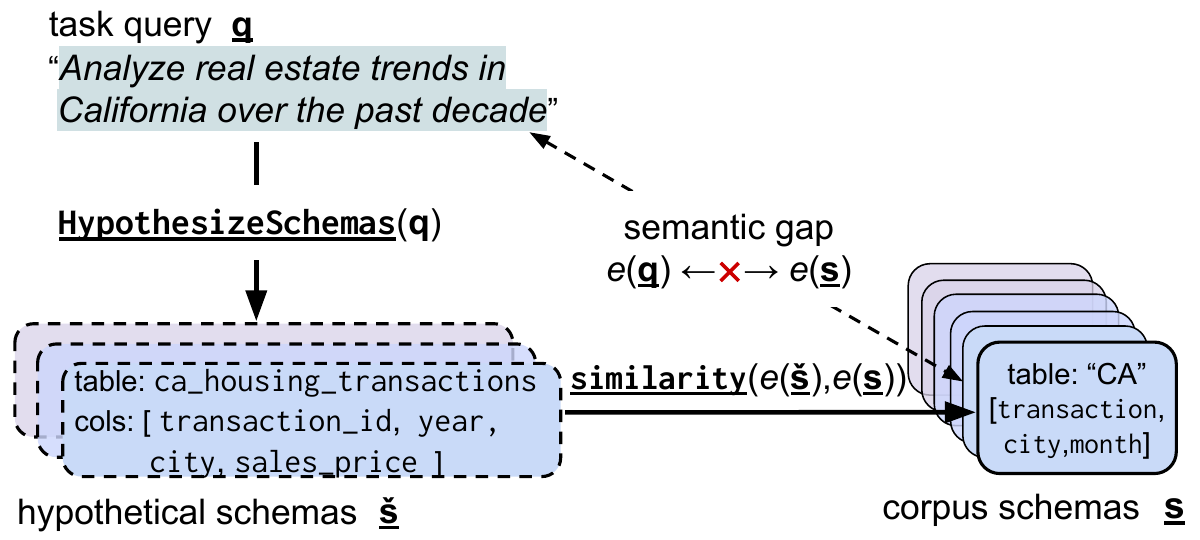}
    \caption{Dataset retrieval through Hypothetical Schema Embeddings (HySE), which uses a language model to hypothesize relevant schemas (\v{s}) for the given task query ({q}). To bridge the semantic gap between input query and schemas in the search corpus, retrieval is based on embeddings of the \textit{hypothetical} schemas and the corpus schemas rather than the input query directly.}
    \label{fig:hyse_fig}
\end{figure}

In the tabular setting, rather than expanding into free-form text, we prompt an LLM to generate hypothetical table schemas--a table name, column names, and two example rows--aligned with the specified task, as illustrated in Figure~\ref{fig:hyse_fig}. We then embed both the task query and the generated schema(s), fuse the embeddings, and perform approximate nearest neighbor (ANN) search over the table representations. This ensures alignment between the input embedding and target representations in the retrieval corpus, while preserving the user’s task specification.

\boldpar{Example.}
Consider the \textit{task query $q$:} ``Study CA housing price trends'' (ln 1--2, Algorithm \ref{alg:hyse}). We first generate $N$ hypothetical schemas for the task query (ln 3-6), embed both the original task query $q$ and the hypothetical schema $J$ (ln 7--8), fuse them with $\lambda{=}0.5$ (ln 9), and perform ANN retrieval over corpus table representations (ln 10). Candidate tables containing overlapping columns such as \texttt{sale\_date}, \texttt{sale\_price}, and geographic attributes (\texttt{city}, \texttt{county}, \texttt{zip\_code}) are thereby promoted, supporting trend analysis across time and location.

\begin{algorithm}[H]
\caption{\textsc{HySE}$(q, N, k)$: task-driven retrieval with hypothetical schemas}
\label{alg:hyse}
\small
\begin{algorithmic}[1]
\State \textbf{inputs}: task query $q$; \# schema samples $N$; fusion weight $\lambda$; return size $k$; LLM $\mathcal{G}$; embedder $E$; ANN index $\mathcal{I}$
\State $S \gets \emptyset$ 
\For{$i \gets 1$ to $N$}
  \State $J_i \gets \mathcal{G}(\textsc{HypothesizeSchema}(q))$ \Comment{generate schema from task}
  \State $S \gets S \cup \{J_i\}$
\EndFor
\State $\mathbf{e}_q \gets E(q)$ \Comment{embed task query}
\State $\mathbf{e}_S \gets \frac{1}{|S|}\sum_{J \in S} E(J)$ \Comment{embed schema(s)}
\State $\tilde{\mathbf{e}} \gets (1-\lambda)\mathbf{e}_q + \lambda\,\mathbf{e}_S$ \Comment{fuse embeddings}
\State \textbf{return} $\textsc{CosSimTopK}_{\mathcal{I}}(\tilde{\mathbf{e}}, k)$ \Comment{return top-$k$ schemas}
\end{algorithmic}
\end{algorithm}

%% file: sections/flexprune.tex
\subsection{FlexPrune: Flexible Metadata Filtering}
\label{sec:nl-refine}

This component addresses the \emph{Hybrid Querying} and \emph{Result Comprehensibility \& Diversity} desiderata (§~\ref{sec:intro}). FlexPrune provides flexibility along two dimensions: users express filtering constraints in natural language rather than through fixed filter interfaces, which typically insufficiently segment the result set to a comprehendible set~\cite{Hulsebos2024HILDA}, allowing for the underlying metadata schema to be extensible rather than limited to predefined filter attributes. While HySE yields semantically aligned candidates, users often need to further constrain search results along metadata dimensions (e.g., granularity and freshness) \cite{Hulsebos2024HILDA}. Traditionally, such attributes are obtained via profiling \cite{Abedjan2015DataProfiling} and surfaced through keyword search or filter widgets \cite{Chapman2019DatasetSearchSurvey}; existing systems explicitly profile spatio-temporal attributes to enable filtering~\cite{Castelo2021Auctus}. However, which metadata dimensions matter is query-dependent \cite{Glavic2024DataValue}, making fixed filter panels a poor fit for diverse information needs.

\begin{figure}%
    \centering
    \includegraphics[width=\linewidth]{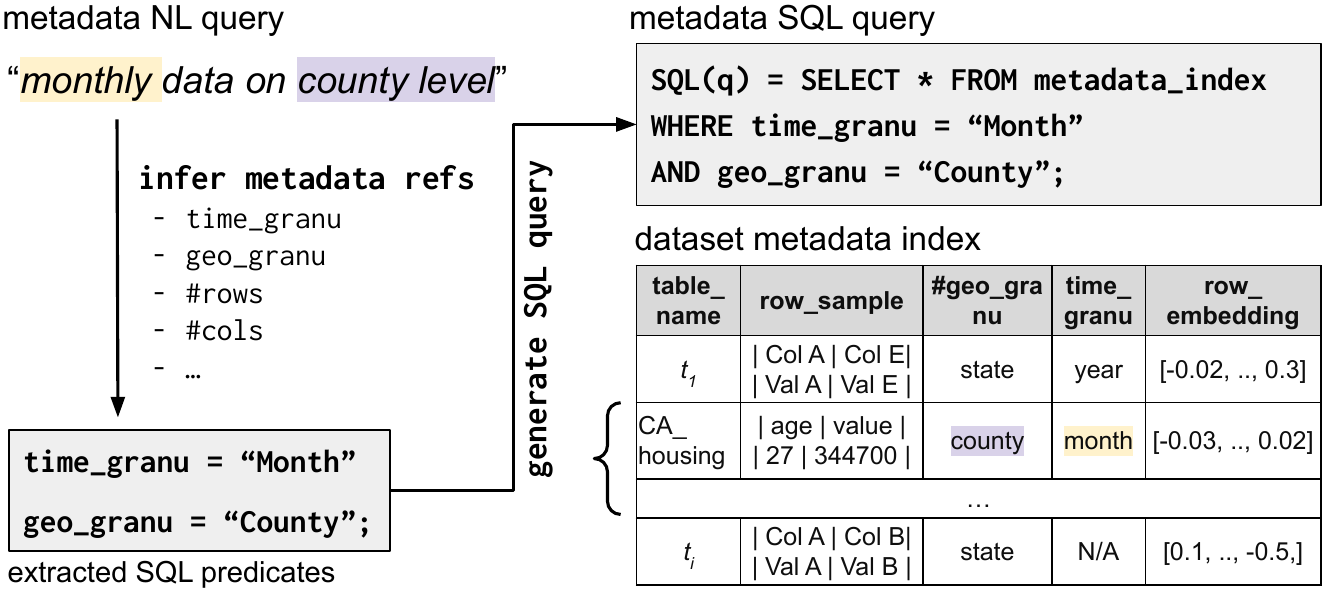}
    \caption{FlexPrune: flexible natural language based metadata filtering using text-to-SQL to enable search space pruning over a customizable metadata index.} %
    \label{fig:flexprune_fig}
\end{figure}

Therefore, FlexPrune resembles an \emph{NL-based metadata refinement} layer: instead of requiring users to navigate predefined filters, the system interprets natural-language constraints (e.g., \emph{``include only monthly data on county level.''}) and compiles them into precise SQL predicates over the hybrid metadata index (§~\ref{sec:offline}), as illustrated in Figure~\ref{fig:flexprune_fig}. This maintains HySE's recall while improving precision and result comprehensibility.

\begin{algorithm}
\caption{\textsc{FlexPrune}$(q, C, k)$: NL-based metadata refinement over hybrid index}
\label{alg:nl-refine}
\small
\begin{algorithmic}[1]
\State \textbf{inputs}: metadata query $q$; current candidate set $C$; return size $k$; field registry $\mathcal{F}$; predicate templates $\Pi$; hybrid metadata table $\mathsf{H}$
\State $M \gets \textsc{InferFields}(q, \mathcal{F})$ \Comment{infer metadata fields in $q$}
\State $\varphi \gets \textsc{BuildSQL}(M, \Pi)$ \Comment{compose SQL predicates}
\State $R \gets \texttt{SELECT*FROM }\mathsf{H}\ \texttt{WHERE table\_name IN } C\ \texttt{AND } \varphi$ \ \ \ \Comment{filter within candidates}
\State \textbf{return} $\textsc{TopK}(R, k)$ \Comment{return top-k matching tables}
\end{algorithmic}
\end{algorithm}

%% file: sections/dataset_construction.tex
\section{\textsc{K\lowercase{aggle}DS} for Task-based Search}
\label{sec:data}

Public benchmarks for \emph{task\,→\,table} retrieval are still missing; we therefore curated a new corpus from Kaggle for training and evaluating such task-driven dataset search systems.\footnote{\url{https://www.kaggle.com/}}
Kaggle is an ideal source because (i) its datasets are explicitly created for data analytics or data science tasks, aligning with our \emph{task–driven} focus, and (ii) its API exposes a uniform JSON metadata schema. We publicly release the resulting corpus as \dataset{}.\footnote{\scriptsize{\url{https://huggingface.co/datasets/trl-lab/KaggleDS}}}

\subsection{Dataset Construction Pipeline}
\label{sec:pre-processing}

\noindent Using the Kaggle CLI\footnote{\texttt{kaggle datasets download}} in May~2024, we extracted \textbf{5,221} dataset ``folders'' (hereafter \emph{datasets}). 
Since our goal is to construct a public benchmark for \textbf{tabular} dataset retrieval, we filter datasets to comprise only CSV files, have a permissive license, and have a non-empty description to ensure input for query inference and column-level metadata enrichment.

\boldpar{Dataset Cardinality} A single Kaggle ``dataset'' often contains several CSV files with related tables instead of just a single CSV file, resulting in the following partitions: 1) single-table: 2,349 datasets contain \emph{exactly one} table, 2) multi-table: 4,154 datasets hold \emph{multiple} tables, of which 1,459 datasets share an identical schema (e.g., yearly partitions), while 2,695 exhibit diverse schemas.

\boldpar{Dataset Splits} To support isolated model training and evaluation, we partition the corpus at the \emph{dataset level} to ensure that no test-set schema occurs in the training data. Of the \textbf{3,141 datasets (6,503 tables)} in total, 60\% 
(1,884 datasets; 4,166 tables) are allocated to a training+validation set, 
with a 2:1 split between training (1,256 datasets; 2,715 tables) and 
validation (628 datasets; 1,451 tables). The remaining 40\% (1,257 datasets; 2,337 tables) are held out as a test set.

\subsection{Metadata Enhancement}
\label{sec:metadata-enhancement}

From Kaggle's metadata and the table content itself, we extract the following metadata: 1) dataset-level metadata: dataset name, description, total downloads (used as a proxy for popularity), usability rating, author-provided tags, and file size. 2) table-level metadata: column headers, example rows, row count, and column count.

The user survey that informs our design criteria~\cite{Hulsebos2024HILDA} also revealed that analysts value a table’s \emph{content granularity}---particularly temporal and geographic coverage, which existing dataset search systems seldom expose. To support such fine-grained filtering, we use GPT-4o to infer two additional metadata fields: \emph{temporal granularity} (Year, Quarter, Month, Week, Day, Hour, Minute, Second) and \emph{geographic granularity} (Continent, Country, State/Province, County/District, City, Zip Code). These attributes are inferred from column headers and example rows using LLM prompts (prompt templates are provided in Appendix~\ref{sec:prompts}).
LLM-generated augmentation has proven effective for overcoming limited query diversity in retrieval settings~\cite{Chen2024ConvAug}.

\subsection{Task Query Construction}

To facilitate evaluation of \emph{task\,→\,table} retrieval through natural language, task-driven, and iterative search queries, we synthesize several query types per table, conditioned on the Kaggle dataset description (see Figure~\ref{fig:kaggle-example-main} for an exemplary instance from the corpus).
\begin{itemize}[leftmargin=*,noitemsep]
  \item \textbf{Task queries:} 3 natural-language search queries based on the dataset description.
  \item \textbf{Keyword queries:} 4 keyword-style phrases for baseline comparison.
  \item \textbf{Metadata queries:} For each task query and metadata field, a corresponding filtering query (e.g., “Include only datasets with more than 1,000 rows”) is generated using GPT-4o, yielding $3 \times N_\text{fields}$ metadata queries per table.
\end{itemize}

This approach follows \citet{Bonifacio2022InParsDataAugmentation}, who show that LLM-generated queries conditioned on document content provide effective supervision for retrieval evaluation. All prompt templates used for generation are provided in Appendix~\ref{sec:prompts}, and a full dataset sample in Appendix~\ref{sec:kaggle-sample} (Figure~\ref{fig:kaggle-example}).

\begin{figure}[t]
\centering
\footnotesize
\setlength{\fboxsep}{6pt}
\fcolorbox{gray!30}{gray!5}{
\begin{minipage}{0.98\columnwidth}
\setlength{\parindent}{0pt}
\setlength{\parskip}{3pt}

\textbf{Dataset metadata:}

\underline{Dataset name:} California house price prediction

\underline{Table name:} \texttt{california\_housing\_test\_1}

\underline{Table header:} \texttt{longitude}, \texttt{latitude}, \texttt{housing\_median\_age}, \texttt{total\_rooms}, \texttt{total\_bedrooms}, \texttt{population}, \texttt{households}, \texttt{median\_income}%

\underline{Description:} The California House Price Prediction system utilizes advanced data analytics to forecast housing prices in the dynamic California real estate market. <.....>  Whether navigating fluctuating market trends or evaluating property investments, this predictive tool serves as a trusted resource for individuals and professionals alike, facilitating strategic and informed housing decisions.\\

\textbf{Queries:}\\
\underline{Task queries:}\\
(1) \textit{Develop a predictive model for California house prices using machine learning algorithms}.\\
(2) \textit{Analyze trends in the California real estate market over the past decade}.

\underline{Metadata queries:}\\
\textit{The dataset should have tags related to artificial intelligence and decision tree; The dataset should have data on at least the city level; The dataset should contain at least 8 columns; The dataset should contain at least 2000 rows}.

\end{minipage}
}

\caption{Truncated example from the \dataset{} corpus showing dataset metadata and generated task- and metadata queries.}
\label{fig:kaggle-example-main}
\vspace{-0.2cm}
\end{figure}

%% file: sections/sys_eval_condensed.tex
\section{Experimental Setup}
\label{sec:eval-setup}

Our evaluation is organized by first evaluating the first retrieval step through \emph{HySE}, followed by a generalization study for open-domain tabular QA. Finally, we assess the end-to-end multi-faceted \emph{pipeline} that chains HySE with FlexPrune for metadata filtering. Here, we elaborate on the full experimental setup.

\subsection{Corpus Index Construction}
\label{sec:offline}

The offline phase transforms each table $t_i \in \mathcal{T}$ into a unified \emph{hybrid metadata index} that integrates symbolic attributes and dense semantic representations.

\boldpar{Metadata ingestion.}
We materialize structural and descriptive attributes extracted from the corpus and enriched via LLM inference (§\ref{sec:metadata-enhancement}). These include temporal and geographic granularity, in addition to dataset-provided metadata.

\boldpar{Semantic serialization and embedding.}
Each table is serialized as a Markdown snippet (\emph{header + sample rows}) and encoded using \texttt{text-embedding-3-small} to produce a 1536-dimensional vector (\texttt{example\_rows\_embed}). 
We adopt Markdown formatting to preserve structural delimiters while keeping the representation lightweight, as LLM understanding of tabular data is sensitive to serialization format~\cite{Sui2024TableMeetsLLM}. This vector is indexed via HNSW~\cite{Malkov2016HNSW} using \texttt{pgvector} to support efficient ANN retrieval.

\boldpar{Hybrid index schema.}
All attributes are stored in a single PostgreSQL table, \texttt{hybrid\_metadata\_index}, enabling both dense similarity retrieval and structured SQL filtering during the online phase. The schema attributes used in our implementation and evaluations comprise: 1) raw metadata fields (number of columns, number of rows, inferred temporal granularity, inferred geographic granularity, and author-provided tags), 2) semantic metadata fields (column names, table name, database name, database-level description, example rows in markdown, and dense embeddings), and 3) auxiliary metadata fields (popularity metrics such as \#downloads, user-reported quality score, file size, LLM-generated keyword queries, task queries, and metadata filters).

This dual representation constitutes a \emph{hybrid index} that combines semantic relevance (via embeddings) and precision-oriented filtering (via symbolic predicates) within a unified layer.

\subsection{Datasets} We use the validation split (4,353 task queries and 1,451 tables, §\ref{sec:data}) of the KaggleDS corpus only for model selection, i.e., to select the HySE configuration (Appendix~\ref{sec:ablation-analysis}).
After identifying the best-performing configuration, all reported results are computed on the held-out test set (7,011 task queries and 2,337 tables, §\ref{sec:data}), ensuring an unbiased assessment of end-to-end performance.

Moreover, we evaluate the HySE retrieval method also on the NQ-Tables dataset~\cite{Herzig2021DTR}, to understand its generalizability (§\ref{sec:eval-nqtables}). NQ-Tables is the table-retrieval benchmark released with DTR: 169,898 Wikipedia tables and 11,628 human-written questions drawn from Natural Questions~\cite{Kwiatkowski2019NQ}, split 9,594 / 1,068 / 966 into train, dev, and test. Its queries are factual (e.g., \textit{``when was the last time easter was in the month of march''}) rather than analytical task descriptions.

\subsection{Metrics} To evaluate the retrieval performance, we adopt the standard metric of recall@$k$, i.e. fraction of queries for which the ground-truth table appears in the top-$k$.

\subsection{HySE Configuration} Unless stated otherwise, HySE prompts \texttt{gpt-4o-mini} with the normalized prompt template (Appendix~\ref{prompt:normalized-hyse-example}) to generate $N{=}2$ hypothetical tables per task query, each comprising a table name, column names, and two example rows. Their mean embedding is fused with the task-query embedding using $\lambda{=}0.5$ (Algorithm~\ref{alg:hyse}), following the proposed approach in HyDE~\cite{Gao2022HyDE}. In line with prior findings of table embeddings for retrieval~\cite{ji2025target,gomm2025metadata}, we encode tables using an embedding of the table name, header, and two example rows (§\ref{sec:offline}). Ablations over schema formulation, number of schemas, table embedding, and query fusion are reported in Appendix~\ref{sec:ablation-analysis}.

\subsection{Baselines}

We compare HySE retrieval against four retrieval baselines spanning syntactic keyword matching, semantic keyword- and task-embedding retriever, and a learned approach, representative of both industry systems and common research abstractions for table retrieval (§~\ref{sec:related-work}).

\noindent\textbf{Syntactic–Keyword Search (BM25).}  A classical search method uses fuzzy BM25 match over \texttt{table\_name} and \texttt{example\_rows\_md}. This reflects common keyword-based search in data lakes or metadata catalogs, and serves as a lower-bound baseline. Queries are matched against pre-tokenized fields while tolerating minor variations.\\
\textbf{Semantic–Keyword Search.}  We embed keywords using \texttt{text-embedding-3-small}, and retrieve tables by cosine similarity against their pre-computed table embeddings. This approach extends keyword search to dense retrieval, enabling semantic matching even when syntactic forms differ (e.g., \textit{“home price”} vs.\ \textit{“housing cost”}).\\
\textbf{Semantic–Task Search.}  The natural-language task query is embedded using \texttt{text-embedding-3-small}. Cosine similarity is computed against pre-computed table embeddings. This baseline represents standard dense retrieval without schema augmentation.\\
\textbf{TAPAS Dense Table Retriever (DTR).}  A \emph{learned} baseline: the TAPAS retriever of \citet{Herzig2021DTR} is fine-tuned on the \dataset{} training split (1,256 datasets; 2,715 tables; 8,142 task queries, §\ref{sec:data}). We initialize from the released ICT-pre-trained dual encoder, encode each table as its title, header, and two example rows to match the corpus representation used by HySE, and train with in-batch negatives following the original recipe. We additionally report \textbf{DTR\,+hn}, the paper's best configuration, which adds one mined hard negative per query.

\subsection{Evaluation Procedure}
All results are reported on the hold-out test split (7,011 task queries over 2,337 tables). For each task query, HySE retrieves from a single corpus-wide embedding index. Recall@$k$ is computed with respect to the ground-truth table.

To analyze structural differences, we partition the evaluation results by properties of the ground-truth table. Results are separately reported for:
(i) \emph{normalized} tables (tables with explicit relational identifiers and separate entities, $n{=}1{,}041$ queries) vs.\ 
(ii) \emph{denormalized} tables (wide tables that collapse multiple entities or attributes into a single schema, $n{=}5{,}970$ queries), and 
(iii) \emph{single-table datasets} ($n{=}2{,}826$ queries) vs.\ 
(iv) \emph{multi-table datasets} ($n{=}4{,}185$ queries) (§~\ref{sec:pre-processing}). 
These segments allow us to assess how schema formulation and dataset cardinality affect retrieval behavior.

\section{Evaluation Results}

Here, we report the results for the three experiments: retrieval with HySE on \dataset{}, its generalization for open-domain tabular question answering, and how FlexPrune complements the search through metadata filtering to condense the result set to avoid occlusion.

\subsection{Retrieval with Hypothetical Schema Embeddings (HySE)}
\label{sec:eval-hyse}

To evaluate HySE, we report \textbf{overall retrieval effectiveness} on the hold-out \textbf{test set}, against both frozen-representation baselines and a retriever trained on our own training split.
Table~\ref{tab:rq1-result-summary} reports Recall@$k$ on the test set under the configuration described in §\ref{sec:eval-setup}.

\begin{table}[H]
\centering
\small
\setlength{\tabcolsep}{6pt}
\caption{HySE retrieval results on \dataset{} test set.}
\label{tab:rq1-result-summary}
\begin{tabular}{@{}lcccc@{}}
\toprule
\textbf{Method} & \textbf{R@10} & \textbf{R@20} & \textbf{R@30}  & \textbf{R@50} \\
\midrule
Syntactic–Keyword & 0.179 & 0.228 & 0.256  & 0.290 \\
Semantic–Keyword  & 0.336 & 0.446 & 0.512  & 0.593 \\
Semantic–Task    & 0.589 & 0.694 & 0.749  & 0.811 \\

\midrule
\multicolumn{5}{@{}l}{TAPAS Dense Table Retrieval (trained)} \\
\addlinespace
\ \ \ \ DTR, no fine-tuning & 0.239 & 0.341 & 0.404  & 0.491 \\
\ \ \ \ DTR & 0.570 & 0.692 & 0.756  & 0.822 \\
\ \ \ \ DTR +hn & 0.571 & 0.687 & 0.751  & 0.820 \\

\midrule
HySE (ours) & \textbf{0.609} & \textbf{0.717}& \textbf{0.773} & \textbf{0.834} \\

\addlinespace
\multicolumn{5}{l}{\textit{Partitioned by Schema Formulation}} \\
\ \ \ \ Normalized & 0.507 & 0.615 & 0.685  & 0.765 \\
\ \ \ \ Denormalized & \textbf{0.627} & \textbf{0.735} & \textbf{0.789} & \textbf{0.846} \\

\addlinespace
\multicolumn{5}{l}{\textit{Partitioned by Dataset Cardinality}} \\
\ \ \ \ Single-table & \textbf{0.777} & \textbf{0.841} & \textbf{0.878} & \textbf{0.913} \\
\ \ \ \ Multi-table & 0.496 & 0.633 & 0.702  & 0.780 \\
\bottomrule
\end{tabular}
\end{table}

\noindent \textbf{Results.} \ \ The performance illustrates that Hypothetical Schema Embeddings enables improvement over direct retrieval using embeddings of the task queries. HySE consistently outperforms semantic-task retrieval (+2.0~pp at R@10; +2.3~pp at R@50), indicating that hypothetical schema generation helps resolve the semantic gap that pure dense embeddings overlook. The comparison against DTR is more demanding: fine-tuning on our training split more than doubles its recall (R@10 $0.239 \rightarrow 0.570$) and lifts it beyond the Semantic--Task baseline at larger cut-offs, yet HySE still leads at every cut-off (+3.9~pp at R@10) \emph{without using any training data}, indicating that our gains do not simply reflect an absence of in-domain supervision in the frozen baselines. We observe that HySE performs particularly well on denormalized tables, while normalized relational-style tables remain more challenging. This suggests that HySE is well-suited to open-data portals and data lakes, where denormalized schemas prevail.
Moreover, retrieval accuracy differs substantially between single-table and multi-table datasets. We observe a particularly high performance of HySE on retrieving single tables where Recall@10 is 0.777 compared to 0.687 by DTR (the single/multi-table results for baselines are suppressed from the table). Across baselines we observe that one-shot multi-table retrieval is challenging, especially for lower values of \textit{k} as correct matches are pushed outside the cut-off, although HySE performs best overall. %
Altogether, our results show that HySE provides strong improvements in recall with low parameter sensitivity and strong generalization across retrieval inputs and corpora, making it practical for table retrieval in typical deployment scenarios.

\subsection{Generalization to Retrieval in Tabular QA}
\label{sec:eval-nqtables}

Evaluating on \dataset{} alone leaves three questions open: whether the results generalize beyond a Kaggle-derived corpus where analytical and predictive tasks are dominant, whether the results hold at a larger corpus scale, and whether HySE's gains are consistent for queries that were not generated. To this end, we evaluate HySE also on NQ-Tables which was introduced for evaluating the TAPAS-based DTR baseline~\cite{Herzig2021DTR}. NQ-Tables is a corpus for Open-Domain Tabular Question Answering comprising 169,898 tables, with real human queries and reference labels.

We report its hold-out test split (959 questions over the full corpus). To demonstrate the robustness of HySE, it is configured \emph{exactly the same} as on \dataset{} (same prompt, schemas $N{=}2$, $\lambda_{\text{task}}=0.5$); the prompt is also deliberately \textbf{not adapted} for answering factual questions, since the claim we test HySE on is that task schema-expansion transfers without per-corpus tuning. For DTR we use the authors' released NQ-Tables retrievers, in the \texttt{large} size their paper reports and in \texttt{medium} for reference, serializing tables as on \dataset{} for comparable evaluation.\\

\noindent\textbf{Results.} \ \ Table~\ref{tab:nqtables} reports the full results. Notably, HySE exceeds DTR-large in all stages up to $+6.0$~pp at R@30, while HySE does not use any NQ-Tables supervision and TAPAS-DTR is fine-tuned on the training split. So, although the evaluation of DTR has a favorable in-domain setting, making this a conservative comparison, these results illustrate the strong benefit of schema expansion with HySE. We interpret this as evidence of generalization of HySE, also beyond data-distribution since NQ-Tables queries are also factual questions rather than analytical task descriptions. As NQ-Tables are not complemented with metadata fields, we cannot use it for evaluating FlexPrune.

\begin{table}[H]
\centering
\small
\setlength{\tabcolsep}{4pt}
\caption{Retrieval performance on the NQ-Tables test split (959 questions, 169,898 tables). %
We reproduce DTR \texttt{large} within one standard deviation of the published results~\cite{Herzig2021DTR}.}
\label{tab:nqtables}
\begin{tabular}{@{}lccccc@{}}
\toprule
\textbf{Method} & \textbf{R@1} & \textbf{R@10} & \textbf{R@20} & \textbf{R@30} & \textbf{R@50} \\
\midrule
DTR +hn (medium) & 0.387 & 0.761 & 0.812 & 0.836  & 0.857 \\
DTR +hn (large) & 0.413 & 0.779 & 0.838 & 0.867  & 0.898 \\
Semantic--Task & 0.368 & 0.819 & 0.890 & 0.922 & 0.944 \\
HySE & \textbf{0.428} & \textbf{0.837} & \textbf{0.897} & \textbf{0.927}  & \textbf{0.955} \\
\bottomrule
\end{tabular}
\end{table}

\subsection{HySE with FlexPrune Metadata Filtering}
\label{sec:eval-pipeline}

Dataset search is typically multi-faceted to reduce occlusion of the result set for a user of the dataset search system, naturally semantic retrieval is therefore followed by a metadata-based filter step. We evaluate the effectiveness of HySE followed by one iteration of flexible metadata-based candidate set filtering with FlexPrune. We construct the two-stage pipeline as follows:
\begin{enumerate}[label=\arabic*),leftmargin=*,noitemsep]
    \item \textbf{HySE Retrieval.}
        Retrieve the top-50 candidates with HySE in the configuration of §\ref{sec:eval-setup}.
    \item \textbf{FlexPrune Filtering.}
        Concatenate metadata subqueries for the task (from fields like \texttt{tags}, \texttt{row\_num}, \texttt{time\_granu}, etc.), and convert the compound prompt into SQL. The resulting query is executed over the metadata schema and intersected with the top-50 HySE candidates (order preserved).
\end{enumerate}

\noindent\textbf{Results.} \ \ Table~\ref{tab:pipeline} shows that a single FlexPrune step applied after HySE retrieval, yields a clear improvement in top-{k} precision. Compared to the baseline search strategy used by existing dataset search engines (semantic query embedding followed by \texttt{tags}-based filtering), we find that the chained pipeline improves \textbf{Recall@10} from $0.644$ to \textbf{0.704}, a gain of \textbf{+6.0 pp} despite only one additional refinement interaction. The improvement is achieved \emph{without} requiring users to enumerate specific columns or understand dataset schemas upfront—HySE provides broad semantic coverage, while metadata filtering sharpens relevance. The recall gain at the top-10 cut-off reduces the number of irrelevant tables a user must inspect. Overall, this chained combination demonstrates that semantic retrieval and symbolic precision are complementary: HySE produces a high-recall candidate set that FlexPrune enables flexible filtering, yielding a practical improvement for dataset search.

\begin{table}[H]
\centering
\setlength{\tabcolsep}{4pt}
\caption{Performance of chained HySE $\rightarrow$ FlexPrune.}
\label{tab:pipeline}
\begin{tabular}{@{}p{5cm}cc@{}}
\toprule
\textbf{System} & \textbf{Recall@10} \\
\midrule
Semantic Task $\rightarrow$ filter-on-tags & 0.644  \\
\addlinespace[2pt]
HySE $\rightarrow$ FlexPrune & \textbf{0.704 ($\Delta$ +6.0)} \\
\bottomrule
\end{tabular}
\end{table}

%% file: sections/conclusion.tex
\section{Conclusion}
\label{sec:conclusion}

Dataset search systems are expected to enable search by task and required metadata. This is typically approached by embedding the input query and performing dense retrieval directly, or training a specialized model to map queries to table schemas followed by metadata search. Dataset search systems should, however, not simply ignore the semantic gap between the user's intent expressed in queries and the targeted schemas, but learned approaches require retraining to the data at hand. We propose a lightweight generalizable mechanism to bridge user-intent through hypothesizing relevant schemas for a given task (HySE), inspired by hypothetical document embeddings. We also introduce an attribute-agnostic mechanism (FlexPrune) for metadata-based filtering that parses metadata queries to SQL, executed on the metadata index at hand. We show on a new large task-based dataset search corpus, \dataset, that HySE substantially improves recall over syntactic and semantic baselines, as well as over a trained dense table retriever while HySE does not require training. We also show that HySE generalizes well to tabular question answering datasets, where it maintains significant performance improvements. When HySE is combined with FlexPrune it increases Recall@10 by +6.0~pp on a held-out test set compared to the traditional approach using direct query embedding and fixed metadata tag search. Altogether, this work shows the potential of lightweight natural-language based mechanisms to enable task-informed tabular dataset search, and provides a stepping stone for future research.

%% file: sections/limitations.tex
\section*{Acknowledgements}
We are grateful for early feedback from Aditya Parameswaran and Shreya Shankar on the ideas presented in this paper. This work was supported by a BIDS-Accenture fellowship, the Dutch Research Council (NWO, AiNed Fellowship grant
NGF.1607.22.045), and an industry grant from SAP. We also acknowledge support through compute credit grants from OpenAI.

\section*{Limitations}
\label{sec:limitation}

\boldpar{Single ground-truth table.} Each query in \dataset{} is associated with exactly one gold table, so Recall@$k$ measures whether that specific table is retrieved. Real analytical tasks are frequently satisfiable by several datasets, which means our metric understates the utility of all methods evaluated here, and possibly unevenly across them. Graded relevance judgments over a pooled candidate set would give a more faithful picture of ranking quality, particularly at small $k$; which we consider future work.

\boldpar{Scale and corpus diversity.} Our primary corpus contains 6{,}503 tables drawn from a single source, Kaggle, whose datasets carry relatively rich author-provided metadata. The NQ-Tables experiment demonstrates that HySE retrieval holds on a corpus of 169{,}898 tables, but we do not report a scaling curve, nor do we evaluate on noisy, sparsely documented tables typical of enterprise data lakes and web-table collections, where metadata-dependent components such as FlexPrune would face materially harder conditions.

%% file: appendix/prompts.tex
\section{LLM Prompts}
\label{sec:prompts}

This appendix provides the complete prompts used in our system.

\vspace{0.2cm}

\newmdenv[
  linewidth=0.5pt, linecolor=gray!50, backgroundcolor=gray!5,
  innerleftmargin=5pt, innerrightmargin=5pt,
  innertopmargin=3pt, innerbottommargin=3pt,
  skipabove=3pt, skipbelow=3pt
]{promptframe}
\newenvironment{promptbox}
  {\begin{promptframe}\scriptsize\setlength{\parskip}{2.5pt}\setlength{\parindent}{0pt}\ignorespaces}
  {\end{promptframe}}

\subsection{Prompts for Metadata Enhancement in Corpus Curation}
\vspace{0.1cm}

\subsubsection{Content Granularity Inference Prompt}
\label{prompt:granularity}

\begin{promptbox}
Given a table with file name "\texttt{\{file\_name\}}" with its header and first few example records: "\texttt{\{table\_markdown\}}", determine the most likely geographic or temporal granularity reflected in the dataset.

Select the temporal granularity from the following options: Year, Quarter, Month, Week, Day, Hour, Minute, or Second. For the geographic granularity, choose from: Continent, Country, State/Province, County/District, City, or Zip Code/Postal Code.

Identify the granularities if reflected based on the table provided, or leave empty if it cannot be inferred from the table.

Generate a dictionary in JSON format with two keys: "\texttt{time\_granularity}" and "\texttt{geo\_granularity}".

Example response format:
\begin{verbatim}
{
    "time_granularity": time_granularity,
    "geo_granularity": geo_granularity
}
\end{verbatim}
\end{promptbox}

\vspace{0.2cm}
\subsection{Prompts for Synthetic Query Generation in Corpus Curation}
\vspace{0.1cm}

\subsubsection{Task Query Generation: Single-Table Databases}
\label{prompt:single-table}

\begin{promptbox}
Based on the dataset description provided below, generate a dictionary in JSON format with the dataset name as key and as values a list of 4 semantically distinct data content keywords that describe the expected content of the dataset but do not describe specific analytic tasks possible with the data. Also, add a list of 3 semantically distinct analytics task sentences that can be performed with the described dataset, e.g. develop ML model to predict XYZ.

If it is hard to complete the task, return an empty dictionary instead.

Dataset Description: "\texttt{\{description\}}"

Example response format:
\begin{verbatim}
{
  "{dataset_name}": {
    "keywords": ["keyword1", "keyword2",
                 "keyword3", "keyword4"],
    "task_queries": ["Task query 1",
                     "Task query 2",
                     "Task query 3"]
  }
}
\end{verbatim}
\end{promptbox}

\subsubsection{Task Query Generation: Multi-Table Databases}
\label{prompt:multi-table}

\begin{promptbox}
Given a dataset that contains CSV files with the file names "\texttt{\{filenames\_str\}}" and the description "\texttt{\{description\}}".

Generate a dictionary in JSON format with the CSV file names as keys and as values a list of 4 semantically distinct data content keywords that describe the expected content of each CSV file but do not describe specific analytic tasks possible with the file.

For each CSV file, also add a list of 3 semantically distinct analytics task sentences that can be performed with the described CSV file, e.g. develop ML model to predict XYZ.

Example response format:
\begin{verbatim}
{
  "csv file 1": {
    "keywords": ["keyword1", "keyword2",
                 "keyword3", "keyword4"],
    "task_queries": ["Task query 1",
                     "Task query 2",
                     "Task query 3"]
  },
  "csv file 2": {
    "keywords": ["keyword1", "keyword2",
                 "keyword3", "keyword4"],
    "task_queries": ["Task query 1",
                     "Task query 2",
                     "Task query 3"]
  }
}
\end{verbatim}
\end{promptbox}

\vspace{0.2cm}

\subsubsection{Metadata Query Generation Prompt}
\label{prompt:metadata-queries}

\begin{promptbox}
Given the analytical task "\texttt{\{task\_query\}}" and the available dataset metadata: "\texttt{\{metadata\}}", generate a list with one natural language metadata query for each metadata attribute that expresses the dataset requirement for that attribute needed to implement the given task.

Output a list of metadata queries that are as specific and concise as possible in the specification. Queries should be within but far from the bound of the given metadata attribute value.

So, with the metadata "Number of Rows" equals to 4,836, the corresponding query could be "The dataset should contain at least 2,000 rows", but could not be "The dataset should contain at least 10000 rows".

For example, for the task "Identify patterns in the trading volume of Adobe's stock" and metadata:
\begin{verbatim}
{
    "Tags": ["Trading", "Forecasting"],
    "Time Granularity": "Day",
    "Number of Columns": 7,
    "Number of Rows": 3,563
}
\end{verbatim}
This could yield: ["The dataset should have the trading tag", "I need a dataset with data on minimum daily level", "The dataset should contain roughly 5 columns", "The dataset should contain at least 2,000 rows"]

Only generate a list of metadata queries, excluding any introductory phrases and focusing exclusively on the tasks themselves.
\end{promptbox}

\vspace{0.2cm}

\subsection{Prompts for LLM Mediation in Online Phase}
\vspace{0.1cm}

\subsubsection{Action Inference Prompt}
\label{prompt:action-infer}

\begin{promptbox}
Given two queries in a search session, decide whether the new query is a "reset" or a "refine" in relation to the previous query.
\begin{itemize}[nosep,leftmargin=1.2em]
\item A "reset" means the new query significantly differs from the previous query, indicating a change in the analytical task focus.
\item A "refine" means the new query builds upon or slightly alters the previous query, indicating a more focused analytical task based on the earlier query.
\end{itemize}
Previous query: "\texttt{\{prev\_query\}}"\\
Current query: "\texttt{\{cur\_query\}}"

Should the search space be reset or refined?
\end{promptbox}

\subsubsection{Field Inference Prompt}
\label{prompt:semantic-fields}

The same prompt template is instantiated separately for two field groups (semantic fields and metadata fields); only the candidate field list differs.

\begin{promptbox}
Given the user's current query, analyze and determine which fields are being referenced. These fields include:

\textit{[Semantic fields]} Table Schema, Example Records, Table Description, Table Tags.

\textit{[Metadata fields]} Table Tags, Number of Columns, Number of Rows, Temporal Granularity, Geographic Granularity.

Current user query: "\texttt{\{cur\_query\}}"

Identify any fields that are explicitly mentioned or strongly implied. Provide your analysis as a structured output listing only those fields that are directly related to the query.
\end{promptbox}

\vspace{0.2cm}
\subsection{Prompts for Hypothetical Schema Generation in Online Phase}
\vspace{0.1cm}

\subsubsection{Normalized Hypothetical Schema Prompt}
\label{prompt:normalized-hyse}

\begin{promptbox}
Given the task of \texttt{\{query\}}, help me generate a normalized database schema to implement the task.

Only generate one table schema, excluding any introductory phrases and focusing exclusively on the tasks themselves.

Generate a JSON with keys as table names and values as column names. For example:

Task: What data is needed to train a machine learning model to forecast demand for medicines across suppliers?

Output:
\begin{verbatim}
{
    "table_name": "Sales",
    "column_names": [
        "sale_id", "medicine_id", "supplier_id",
        "quantity_sold", "sale_date", "price", "region"
    ]
}
\end{verbatim}
\end{promptbox}

\vspace{0.2cm}

\subsubsection{Denormalized Hypothetical Schema Prompt}
\label{prompt:denormalized-hyse}

\begin{promptbox}
Given the task of \texttt{\{query\}}, help me generate a denormalized database schema to implement the task.

Only generate one table schema, excluding any introductory phrases and focusing exclusively on the tasks themselves.

Generate a JSON with keys as table names and values as column names. For example:

Task: What data is needed to train a machine learning model to forecast demand for medicines across suppliers?

Output:
\begin{verbatim}
{
    "table_name": "Medicine_Sales",
    "column_names": [
        "medicine_name", "supplier_name", "quantity_sold",
        "sale_date", "price", "region"
    ]
}
\end{verbatim}
\end{promptbox}

\vspace{0.1cm}

\subsubsection{Normalized Hypothetical Schema Prompt (with Example Rows)}
\label{prompt:normalized-hyse-example}

\begin{promptbox}
Given the task: "\texttt{\{query\}}"

Generate a normalized database schema specifically designed for this task. Think carefully about what data would be needed to accomplish this specific goal.

Requirements:
\begin{enumerate}[label=\arabic*.,leftmargin=15pt,nosep]
\item Create ONE table that directly supports the given task
\item Use normalized relational design principles (separate entities, use IDs for relationships)
\item Column names should reflect the specific domain and requirements of the task
\item Table name should be descriptive and task-relevant
\item Provide realistic example data that matches the task context
\end{enumerate}

Generate a JSON with this exact structure:
\begin{verbatim}
{
    "table_name": "your_task_specific_table_name",
    "column_names": ["col1", "col2", "col3", ...],
    "example_2rows": [
        [val1, val2, val3, ...],
        [val1, val2, val3, ...]
    ],
    "example_3rows": [
        [val1, val2, val3, ...],
        [val1, val2, val3, ...],
        [val1, val2, val3, ...]
    ]
}
\end{verbatim}

Focus on the specific requirements of "\texttt{\{query\}}" - what entities, attributes, and relationships would be needed?
\end{promptbox}

\subsubsection{Denormalized Hypothetical Schema Prompt (with Example Rows)}
\label{prompt:denormalized-hyse-example}

\begin{promptbox}
Given the task: "\texttt{\{query\}}"

Generate a denormalized database schema specifically designed for this task. Think carefully about what data would be needed to accomplish this specific goal.

Requirements:
\begin{enumerate}[label=\arabic*.,leftmargin=15pt,nosep]
\item Create ONE denormalized table that captures all necessary information for the task
\item Use descriptive, human-readable column names (avoid IDs, prefer natural keys)
\item Include all relevant attributes directly in the table (no foreign key relationships)
\item Column names should reflect the specific domain and requirements of the task
\item Table name should be descriptive and task-relevant
\item Provide realistic example data that matches the task context
\end{enumerate}

Generate a JSON with this exact structure:
\begin{verbatim}
{
    "table_name": "your_task_specific_table_name",
    "column_names": ["descriptive_col1", "descriptive_col2",
        "descriptive_col3", ...],
    "example_2rows": [
        [val1, val2, val3, ...],
        [val1, val2, val3, ...]
    ],
    "example_3rows": [
        [val1, val2, val3, ...],
        [val1, val2, val3, ...],
        [val1, val2, val3, ...]
    ]
}
\end{verbatim}

Focus on the specific requirements of "\texttt{\{query\}}" - what information would be needed in a single, comprehensive table?
\end{promptbox}

\subsection{Prompts for NL-to-SQL Metadata Predicate Construction in Online Phase}

\subsubsection{NL-to-SQL Metadata Predicate Prompt}
\label{prompt:nl-to-sql}

\begin{promptbox}
Given these requirements, convert the identified metadata fields and the user query into precise SQL WHERE clause conditions. Identified fields: \texttt{\{identified\_fields\}}; current user query: "\texttt{\{cur\_query\}}". Follow these rules:
\begin{enumerate}[label=\arabic*.,leftmargin=15pt,nosep]
\item Temporal Granularity: must be exactly one of Year, Quarter, Month, Week, Day, Hour, Minute, Second.
\item Geographic Granularity: must be exactly one of Continent, Country, State/Province, County/District, City, Zip Code/Postal Code.
\item Numeric Columns (e.g., \texttt{column\_numbers}, \texttt{row\_numbers}): use operators like $\geq$, =, etc., based on user query context (e.g., "at least 5 columns" $\Rightarrow$ "$\geq$ 5").
\item Tags (\texttt{table\_tags}): never combine multiple tags in a single line such as \texttt{LIKE '\%business\%' AND table\_tags LIKE '\%education\%'}. Instead, if the user wants ``business and education'' tags, produce two separate lines:
\begin{verbatim}
{ "field": "table_tags", "clause": "= 'business'" },
{ "field": "table_tags", "clause": "= 'education'" }
\end{verbatim}
letting the final code treat them as AND (or OR) conditions as needed.
\item Output in JSON with a top-level key \texttt{sql\_clauses} holding an array of objects. Each object should have ``\texttt{field}'' and ``\texttt{clause}''. Example:
\begin{verbatim}
{
  "sql_clauses": [
    { "field": "table_tags", "clause": "= 'finance'" },
    { "field": "temporal_granu", "clause": "= 'day'" },
    { "field": "row_numbers", "clause": ">= 3000" }
  ]
}
\end{verbatim}
\item No Combined Clause: do not produce lines like \texttt{table\_tags LIKE '\%tag1\%' AND table\_tags LIKE '\%tag2\%'}. Always split multi-tag queries into multiple items in the \texttt{sql\_clauses} array.
\end{enumerate}

Make sure your final output is valid JSON, with no extra keys, so each field can be parsed separately. Now, please produce the final structured SQL clauses:
\end{promptbox}

%% file: appendix/kaggle_example.tex
\section{Kaggle Dataset Sample}
\label{sec:kaggle-sample}

Figure~\ref{fig:kaggle-example} shows a complete \dataset{} instance: the dataset metadata, all three generated task queries with their metadata queries, and the keyword queries used by the baselines.

\begin{figure}[h!]
\centering
\footnotesize
\setlength{\fboxsep}{6pt}
\fcolorbox{gray!30}{gray!5}{
\begin{minipage}{0.98\columnwidth}
\setlength{\parindent}{0pt}
\setlength{\parskip}{3pt}

\textbf{Dataset metadata:}

\underline{dataset name:} California house price prediction

\underline{Table name:} \texttt{california\_housing\_test\_1}

\textit{Table header:} \texttt{longitude}, \texttt{latitude}, \texttt{housing\_median\_age}, \texttt{total\_rooms}, \texttt{total\_bedrooms}, \texttt{population}, \texttt{households}, \texttt{median\_income}, \texttt{median\_house\_value}, \texttt{ocean\_proximity}

\underline{Description:} The California House Price Prediction system utilizes advanced data analytics to forecast housing prices in the dynamic California real estate market. Drawing from a diverse range of reputable sources including real estate listings, property datasets, and government records, this system ensures a robust foundation for analysis. Through meticulous data collection, cleaning, and feature engineering processes, relevant attributes such as property specifications, historical sales data, and neighborhood characteristics are carefully curated to enhance predictive accuracy. Powered by machine learning algorithms, the system provides stakeholders with invaluable insights, empowering them to make informed decisions regarding buying, selling, or investing in California real estate. Whether navigating fluctuating market trends or evaluating property investments, this predictive tool serves as a trusted resource for individuals and professionals alike, facilitating strategic and informed housing decisions.

\textbf{Queries:}\\
\underline{Task queries:}\\
(1) \textit{Develop a predictive model for California house prices using machine learning algorithms}.\\
(2) \textit{Analyze trends in the California real estate market over the past decade}.\\
(3) \textit{Evaluate the impact of neighborhood characteristics on property prices in California}.

\underline{Metadata queries:}\\
(1) \textit{The dataset should have tags related to artificial intelligence and decision tree; The dataset should have data on at least the city level; The dataset should contain at least 8 columns; The dataset should contain at least 2000 rows}.\\
(2) \textit{The dataset should contain categorical and decision tree tags; The dataset should provide city-level data for California; The dataset should contain at least 5 columns; The dataset should contain at least 2000 rows}.\\
(3) \textit{The dataset should have tags related to social issues and advocacy; The dataset should have city-level geographic granularity for California; The dataset should contain approximately 11 columns; The dataset should contain at least 2500 rows}.

\textbf{Baseline queries:}\\
\underline{Keyword queries:} real estate listings; property specifications; historical sales data; neighborhood characteristics
\end{minipage}
}
\caption{Example from the \dataset{} corpus showing (1) dataset metadata, (2) task and metadata queries, and (3) keyword queries as used in baselines.}
\label{fig:kaggle-example}
\end{figure}

%% file: appendix/ablation_analysis.tex
\section{Ablation Analysis}
\label{sec:ablation-analysis}

\subsection{Ablation Experiment} We use the \textbf{validation set} to analyze various operational configurations for HySE via targeted ablations over settings for the schema formulation, schema count, and corpus encoding. Table~\ref{tab:hyse-ablations} shows the results for three ablation analyses.

\begin{table}[H]
\centering
\scriptsize
\setlength{\tabcolsep}{3.8pt}
\caption{HySE ablation studies on the validation set. Each block varies one design dimension of HySE while holding the others at the selected configuration. Baselines are shared across ablations and reported once for reference.}
\label{tab:hyse-ablations}
\begin{tabular}{llccccc}
    \toprule
    & & \textbf{R@10} & \textbf{R@20} & \textbf{R@30} & \textbf{R@50} \\
    \midrule
    
    \multicolumn{6}{l}{\textbf{Baselines}}\\
    \quad Semantic--Task     &  & .546 & .624 & .679  & .760 \\
    \quad Semantic--Keyword  &  & .338 & .432 & .500  & .595 \\
    \quad Syntactic--Keyword &  & .177 & .218 & .239  & .272 \\
    \midrule
    
    \multicolumn{6}{l}{\textbf{Schema Formulation}}\\
    \quad Normalized     &  & \textbf{.566} & \textbf{.645} & \textbf{.701} & \textbf{.786} \\
    \quad Denormalized   &  & .564 & .640 & \textbf{.701} & .782 \\
    
    \multicolumn{6}{l}{\textbf{Schema Count ($N$)}}\\
    \quad $N{=}1$ &  & .563 & .645 & \textbf{.702}  & .783 \\
    \quad $N{=}2$ &  & \textbf{.566} & .645 & .701  & \textbf{.786} \\
    \quad $N{=}3$ &  & .562 & \textbf{.646} & .700  & .785 \\
    
    \multicolumn{6}{l}{\textbf{Corpus Encoding}}\\
    \quad C1: Header only             & & .433 & .515 & .572 & .647 \\
    \quad C2: Header+2 rows           & & .507 & .591 & .645  & .722 \\
    \quad C3: Header+3 rows           & & .510 & .589 & .643  & .723 \\
    \quad C4: Name+Header        & & \textbf{.566} & .641 & .688  & .756 \\
    \quad C5: Name+Header+2 rows      & & \textbf{.566} & \textbf{.645} & \textbf{.701}  & \textbf{.786} \\
    \quad C6: Name+Header+3 rows      & & .561 & .641 & .697  & .780 \\
    
    \bottomrule
    \end{tabular}
\end{table}

\textbf{Key insights from ablations.}
We observe that relationally structured (normalized) schemas yield the highest recall across $k$; performance is largely insensitive to the number of generated schemas, with $N \in \{1,2\}$ offering strong results at minimal compute overhead; and finally ``table name + header + two rows'' provides the most effective representation of candidate tables.

While the validation favors normalized prompts, HySE attains its strongest test-set performance on \emph{denormalized} tables, which dominate open-data corpora. We hypothesize this is because denormalized tables expose richer explicit semantic signals in their wide headers and example rows; HySE’s hypothetical schemas align well with such semantic attributes, and joint fusion preserves task phrasing while adding structural cues. This robustness to heterogeneous schema styles is encouraging for open-data repositories where schema standardization is rare.

\vspace{0.5em}
\emph{HySE Variation Experiment.}
Following the HyDE paradigm~\cite{Gao2022HyDE}, HySE can operate in two embedding modes:

\begin{itemize}[leftmargin=*]
  \item \textbf{Schema-only embedding:}  
        Compute similarity using only the vector(s) from the LLM-generated hypothetical schema(s). This models the retrieval process as fully driven by task-imagined schema content.
  \item \textbf{Joint fusion embedding:}  
        Average the task query embedding with the schema embedding(s). This retains the original user intent in the dense vector, potentially grounding ambiguous schema generations.

        We use $\lambda{=}0.5$ (Algorithm~\ref{alg:hyse}), i.e., equal weight on the task and schema embeddings. Both variants use $N{=}2$ normalized schemas and retrieve against the table name + header + 2 example rows corpus (C5).
\end{itemize}

\begin{table}[H]
\centering\small
\caption{Effect of query fusion on retrieval ($N{=}2$, normalized schemas, table name + header + 2 example rows corpus).}
\label{tab:rq4-query-fusion}
\begin{tabular}{lcccc}
\toprule
\textbf{Embedding Strategy} & R@10 & R@20 & R@30 & R@50 \\
\midrule
HySE – Schema-only   & .552 & .638 & .694 & .774 \\
HySE – Joint fusion  & \textbf{.566} & \textbf{.645} & \textbf{.701} & \textbf{.786} \\
\midrule
Semantic–Task     & .546 & .624 & .679 & .760 \\
Semantic–Keyword  & .338 & .432 & .500 & .595 \\
Syntactic–Keyword & .177 & .218 & .239 & .272 \\
\bottomrule
\end{tabular}
\end{table}

\vspace{0.5em}
\textbf{Results.}
Table~\ref{tab:rq4-query-fusion} compares schema-only and joint fusion embeddings. We make the following observations.

\textbf{Joint fusion yields the best overall performance.} Combining the task query embedding with hypothetical schema embeddings improves recall across all cut-offs, with gains of +1.4 pp at R@10 and +1.2 pp at R@50 over schema-only.

\textbf{Schema-only remains highly competitive.} Even without incorporating the task query itself, HySE surpasses the Semantic–Task baseline by +0.6–1.5 pp across R@10–50, indicating that generated schema structure alone provides strong retrieval alignment.

\textbf{Gains are most helpful at early ranks.}  The largest improvement from fusion is at R@10, suggesting that the task query helps sharpen top-ranked disambiguation by preserving user intent alongside the structural signal.

\textbf{Fuse when top-$k$ matters.}  In interactive settings where users depend heavily on the first page of results, joint fusion offers a small but meaningful quality boost with negligible added cost.